\documentclass[10pt, onecolumn, final]{elsart}
\usepackage[utf8]{inputenc}

\usepackage{natbib}
\usepackage{graphicx}
\usepackage{amsmath}
\usepackage{amssymb}
\usepackage{url}

 \journal{Earth and Planetary Science Letters}

\begin{document}

\begin{frontmatter}
  \title{Temporal Variations of Strength and Location of the South
    Atlantic Anomaly as Measured by \textsl{RXTE}}
  
  \author[rem,ecap]{Felix~F\"urst\corauthref{cor}}
  \corauth[cor]{Corresponding author.}
  \ead{felix.fuerst@sternwarte.uni-erlangen.de}
  \author[rem,ecap]{J\"orn~Wilms}
  \ead{joern.wilms@sternwarte.uni-erlangen.de}
  \author[cass]{Richard~E.~Rothschild}
  \ead{rrothschild@ucsd.edu}
  \author[cass,gsfc,cresst]{Katja~Pottschmidt}
  \ead{katja@milkyway.gsfc.nasa.gov}
  \author[ucsc]{David~M.~Smith}
  \ead{dsmith@scipp.ucsc.edu}
  \author[cass]{and Richard~Lingenfelter}
  \ead{rlingenfelter@ucsd.edu}

  \address[rem]{Dr.\ Karl-Remeis-Sternwarte, Sternwartstr.\ 7, 96049
    Bamberg, Germany}
  \address[ecap]{Erlangen Centre for Astroparticle Physics,
    Erwin-Rommel-Stra\ss{}e 1, 91058 Erlangen, Germany} 
  \address[cass]{Center for Astrophysics and Space Sciences, University of
    California at San Diego, La Jolla, 9500 Gilman Drive, CA 92093-0424,
    USA}
  \address[gsfc]{CRESST and NASA Goddard Space Flight Center, Astrophysics Science
    Division, Code 661, Greenbelt, MD 20771, USA}
  \address[cresst]{Center for Space Science and Technology, University of Maryland Baltimore County, 1000
    Hilltop Circle, Baltimore, MD 21250, USA}
  \address[ucsc]{SCIPP, University of California at Santa Cruz, 1156 High Street, Santa Cruz, CA 95064, USA}

  \begin{abstract}
    The evolution of the particle background at an altitude of
    $\sim$540\,km during the time interval between 1996 and 2007 is
    studied using the particle monitor of the High Energy X-ray Timing
    Experiment on board NASA's Rossi X-ray Timing Explorer. A special
    emphasis of this study is the location and strength of the South
    Atlantic Anomaly (SAA). The size and strength of the SAA are
    anti-correlated with the the 10.7\,cm radio flux of the Sun, which
    leads the SAA strength by $\sim$1\,year reflecting variations in solar heating of the upper atmosphere. The location of the SAA is also found to
    drift westwards with an average drift rate of about 0.3$^\circ / \text{yr}$
 following the drift of the geomagnetic field configuration. Superimposed to this drift
    rate are irregularities, where the SAA suddenly moves eastwards
    and where furthermore the speed of the drift changes. The most
    prominent of these irregularities is found in the second quarter
    of 2003 and another event took place in 1999. We suggest that these events are previously unrecognized manifestations of the geomagnetic jerks of the Earth's magnetic field.
  \end{abstract}
  
  
 \end{frontmatter}

\section{Introduction}\label{sec:intro}
Geomagnetically trapped ionized particles, mainly electrons and
protons, are a hazard to modern spaceflight. They can cause failure of
microprocessors aboard satellites, bias measurements of cosmic
sources, and pose a health hazard to astronauts. In low-Earth orbits
(LEO), at a height of several 100\,km above the Earth's surface, a space
vehicle encounters the most intense radiation in a region called the
South Atlantic Anomaly \citep[SAA;][]{kurnosova62a}. In this region
above the South Atlantic just off the Brazilian coast, the level of
ionizing radiation can be increased by several orders of magnitude.
This increase can be explained by a weakened magnetic field, which
causes the geomagnetically trapped particles to mirror at lower
altitudes, increasing the local particle flux \citep[see,
e.g.,][]{evans02a}. While the SAA is an extended region, it is common
to define its position as the 2D-locus of the maximum of the local
particle flux at a given altitude \citep{heynderickx96a}.

\subsection{The Earth's magnetic field}
To a first crude approximation, the Earth's magnetic field can be
modeled as a dipole with an axis that is is offset from the Earth's
center by about 500\,km towards Southeast Asia and inclined by about
11\,degrees with respect to the Earth's rotational axis
\citep[see][]{pinto91a}. At a given altitude above the Earth's surface,
in this approximation the weakest $B$-field is above the South
Atlantic, the so-called South Atlantic Magnetic Anomaly (SAMA). While it is not a priori required that
the local maximum of ionizing flux is located close to the minimum of
the magnetic field, as its location also depends on the particle
distribution in the atmosphere \citep{lauriente96a}, there is a good
correlation between the maximum of the ionizing flux and the $B$-field. 
The primary physical cause for the Earth's $B$-field is, however, dynamic currents in the outer core. These currents, consisting mainly of molten iron, produce a self-sustaining dynamo which gives rise to the geomagnetic-field. The ``Dynamo Theory'' was first proposed by \citet{elsasser46a} to be applicable to the Earth. See \citet{roberts00a} for a review and a discussion of the stability of the dynamo. Despite the overall stability of the dynamo, the field configuration is slowly changing, giving rise to the secular variation of the field \citep[see][for a short historical review]{malin85a}. From archeomagnetic data \citet{pinto91a} have shown that the SAMA has been slowly drifting westwards for at least 1400\,years. This long-term drifting was recently confirmed by \citet{mandea07a}, using data calculated via the GUFM model \citep{jackson00a} and the CHAOS model \citep[model from the satellites \textit{CHAMP}, \textit{\O{}rsted} and \textit{SAC-C},][]{olsen06a}.  
From time to time, however, an abrupt change in the secular
variation is measured, which is called a ``geomagnetic jerk''. These
jerks occur when the fluid flows on the core surface change. Due to the complex nature of the geodynamo, such jerks are rather common. Jerks are found in geomagnetic data of the last century and show up as changes
in the time derivative of the eastward-pointing component of the magnetic field.
See \citet{wardinski08a} and references therein for a description of
the underlying physical processes.

\subsection{The particle environment}
In agreement with the SAMA location the maximum of the particle flux (SAA) is moving westwards as well.
Because the aforementioned risks of high radiation to modern
spaceflight make it necessary to know the location and strength of the
SAA at any given time, a large number of measurements and
investigations of this behavior have been carried out over the years. The constant
change of the position of the SAA requires to monitor constantly the space environment and to develop models which predict the
future movement of the SAA. Most of the previous works on the movement
of the SAA compare recent measurements of the particle flux to the
particle flux predicted by the \textit{AP-8}/\textit{AE-8} models. The \textit{AP-8} model for protons was released in
1976 and includes all data taken after 1970 \citep{sawyer76a}, while the \textit{AE-8}
for electrons was released  in 1991 \citep{vette91a}. Both models together include data
from over 43 satellites \citep{barth03a}. Nonetheless, these models
are based on data taken with older instruments which were not
necessarily as sophisticated as modern
devices. With the help of these models, \citet{konradi94a} and
\citet{badhwar94a} have shown that the SAA is located distinctly more
westwards in modern measurements, leading to an average annual
movement of about $0.3^\circ / \text{yr}$. The derived
drift rate depends strongly on the accuracy of the models.
\citet{grigoryan05a} have shown that the SAA has somewhat different
sizes and locations in different energy bands. 
This change of shape leads to a systematic error when comparing older
models to recent measurements, which have been made using different
detectors. In an effort to compare measurements made with an identical
setup, \citet{badhwar97a} compared data taken 21.2\,years apart, on
\textit{Skylab} during December 1973 and January 1974 and on
\textit{Mir} in March 1995. He found a drift rate of $0.28
\pm 0.03^\circ / \text{yr}$ westward which is in very good
agreement with other measurements.

\citet{ginet06a} used a similar approach in comparing measurements
taken between 2000 and 2006 with the Tri-Service Experiments Mission-5
(\textit{TSX-5}) to 1994--1996 data from the Advanced Photovoltaic and
Electronic Experiments (\textit{APEX}), resulting in a somewhat faster
movement of $0.43 \pm 0.13^\circ / \text{yr}$. The long
average of the \textit{TSX-5} mission might be responsible for that
discrepancy \citep{ginet06a}. Another indirect way to measure the
particle flux is to record Single Event Upsets (SEU) or Single Event
Effects (SEE) in electronic devices, see, e.g., \citet{lauriente96a},
\citet{mullen98a} or \citet{adolphsen95a}. These measurements bear a
higher systematic error, as the upsets can be caused by particles with
different energies or even by cosmic rays. This uncertainty is
reflected in the larger scatter of the derived movement rates, which
vary between $0.19^\circ / \text{yr}$ and
$0.4^\circ / \text{yr}$.

\subsection{Long time measurement with \textit{RXTE}}
A shortcoming of all of these previous studies, however, is that since
most of the measurements were taken with detectors on missions with a
short lifetime, such as manned spacecraft, only pointwise data are
available, even if they were made over a long range of time. This
means that short time variations of the drift rate of the SAA had to
be ignored. In this paper we study the SAA based on data from the
\textit{Rossi X-ray Timing Explorer} (\textit{RXTE}), taken
continuously between 1996 and 2007. This is the first time that the SAA is studied on such a long and continuous data basis. In Sect.~\ref{sec:withRXTE} we
describe our detection method and our data extraction pipeline. In
Sect.~\ref{sec:results} our findings about the SAA are
presented, with Sect.~\ref{susec:tempvar} focusing on the temporal evolution of the strength of the SAA and Sect.~\ref{susec:move} concentrating on the location and the drift of the SAA. We summarize our results
and discuss some future investigations in Sect.~\ref{sec:disc}.

\section{Instrumentation and data acquisition}\label{sec:withRXTE}
\subsection{The Rossi X-ray Timing Explorer}\label{sec:rxte}
\begin{figure}\centering
  \includegraphics[viewport=56 363 554 738, clip, width=0.7\columnwidth]{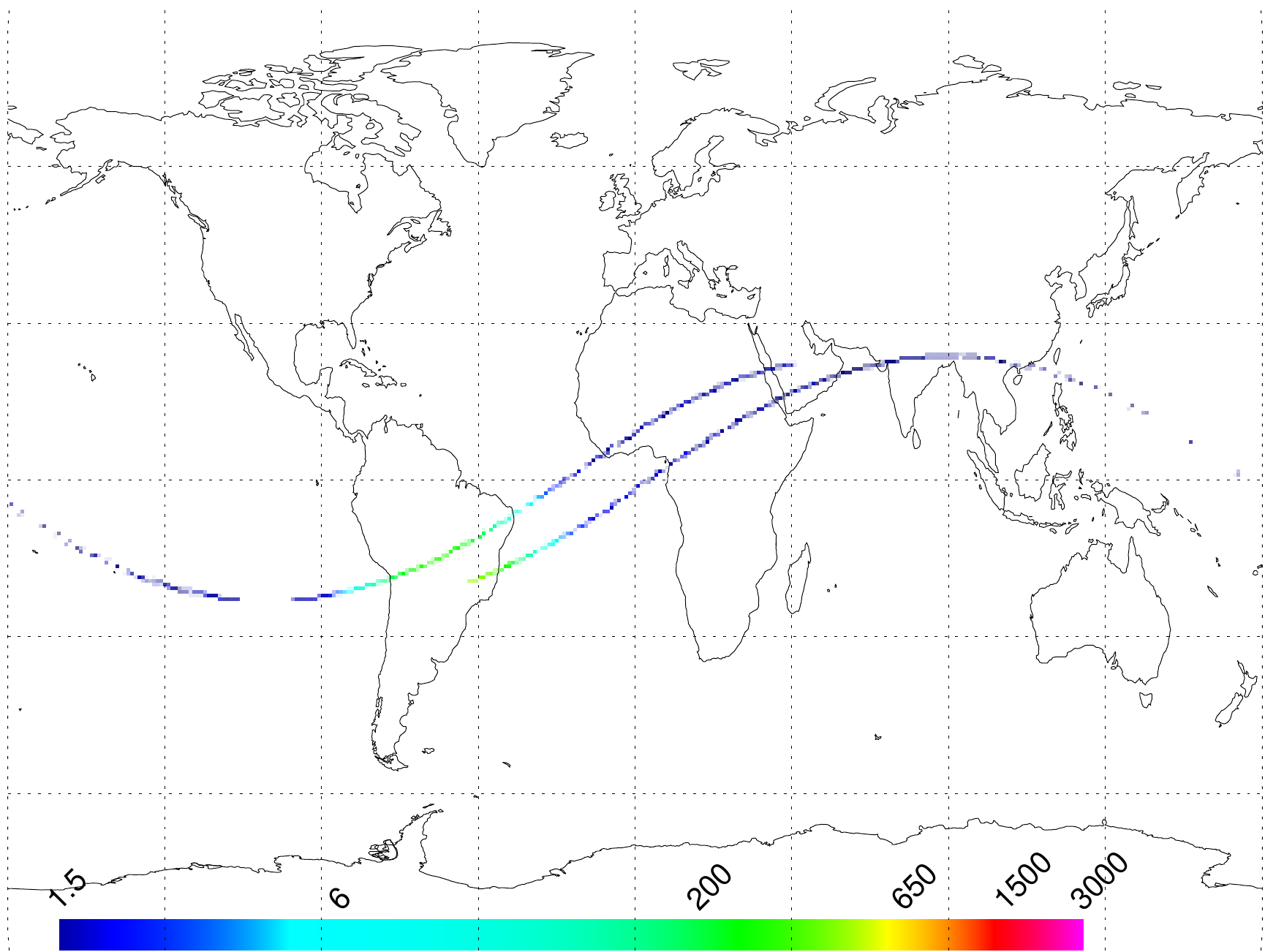}
 \vspace{1cm}
  \caption{Orbit and particle monitor rate of \textit{RXTE} on 26 June 2005 from 00:06\,h to 02:14\,h in the altitude bin 490--499\,km, showing 1$\frac 1 4$ orbits. The scarce datapoints above the Pacific are due to the aforementioned orbital perturbations.}
  \label{fig:shorbit}
\end{figure}

The presented analysis is based on data taken with instruments on the
Rossi X-ray Timing Explorer (\textit{RXTE}), a satellite devoted to
X-ray astronomy. \textit{RXTE} was launched on 1995 December 30 into a
quasi-spherical orbit at an altitude of 592\,km above the Earth. The
orbital inclination is $23^\circ$ and the orbital period is 90\,min, such that the satellite passes directly through the SAA about 8 times a day on average. See Fig. \ref{fig:shorbit} for a plot of the orbit of \textit{RXTE} on June 26 2005 from 00:06\,h to 02:14\,h.
The slight eccentricity of $e = 0.0008$ and orbital perturbations result in an altitude
 interval of $\sim$30\,km being covered by the satellite per orbit.  Due to atmospheric drag, the
satellite's altitude is slowly decreasing, reaching 488\,km in
December 2007. 

\begin{figure}\centering
  \includegraphics[width=0.45\textwidth]{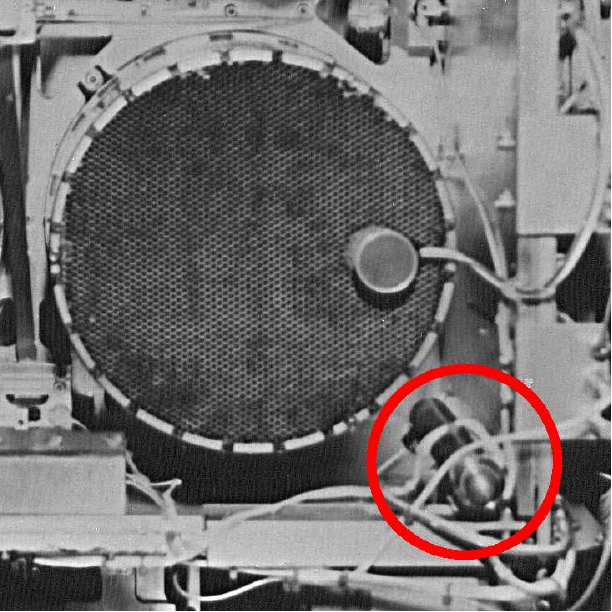}
 \vspace{1cm}
  \caption{Photograph of portion of a HEXTE cluster showing the collimator above one of four phoswich scintillation detectors. The particle monitor is located in the
    red circle in the lower right.}
  \label{fig:hexte-pm}
\end{figure}

One of the three instruments on board \textit{RXTE} is the High Energy
X-ray Timing experiment\footnote{\url{http://heasarc.gsfc.nasa.gov/docs/xte/HEXTE.html}} \citep[HEXTE;][]{rothschild98a}. HEXTE
consists of two independent clusters of four NaI(Tl)/CsI(Na) phoswich
scintillation counters, called HEXTE~A and HEXTE~B. The
photomultiplier tubes attached to the scintillation counters are
operated at a voltage of 845\,V. As the photomultipliers can easily be
damaged by exposure to energetic particles, the
high-voltage is reduced to 220\,V when passing through regions of
increased background flux. In order to do so, the background in each
cluster is constantly monitored using particle monitors consisting of
a 1.27\,cm diameter by 1.27\,cm thick plastic scintillator cylinder
and a 0.5\,inch photomultiplier tube. One particle monitor is attached
to each cluster (Fig.~\ref{fig:hexte-pm}). The monitors are shielded
by aluminum, providing a threshold energy of around 0.8\,MeV for
electrons and 17\,MeV for protons. Due to slight differences in the radiation environment of
the satellite, while the count rates from the two monitors show very
similar behavior, HEXTE~A's rates are typically 2\% larger than
HEXTE~B's rates (Fig.~\ref{fig:simult_fit}). In the later analysis,
this offset was taken into account by introducing a multiplicative
flux normalization constant in all fits and keeping all other
parameters describing the shape of the SAA the same for both
detectors.

Both monitors remain fully functional throughout the SAA passages. The
count rate in each of the two particle background monitors is taken
with a time resolution of 16\,s and is available in the housekeeping data
which are part of the standard \textsl{RXTE} data products. In this
paper data taken between January 1996 and January 2008 are taken into
account, corresponding to $\sim$$23\times 10^6$ data points in
$\sim$20100 individual housekeeping data files for each cluster. We
therefore have detailed and quasi-uninterrupted background radiation
measurements at \textsl{RXTE}'s altitude available for a time interval
of 12\,years. In terms of time coverage, these data represent one of
the best statistics of the SAA region investigated up to date.

\begin{figure*}
   \includegraphics[viewport=3 3 502 375, clip, width=0.7\columnwidth]{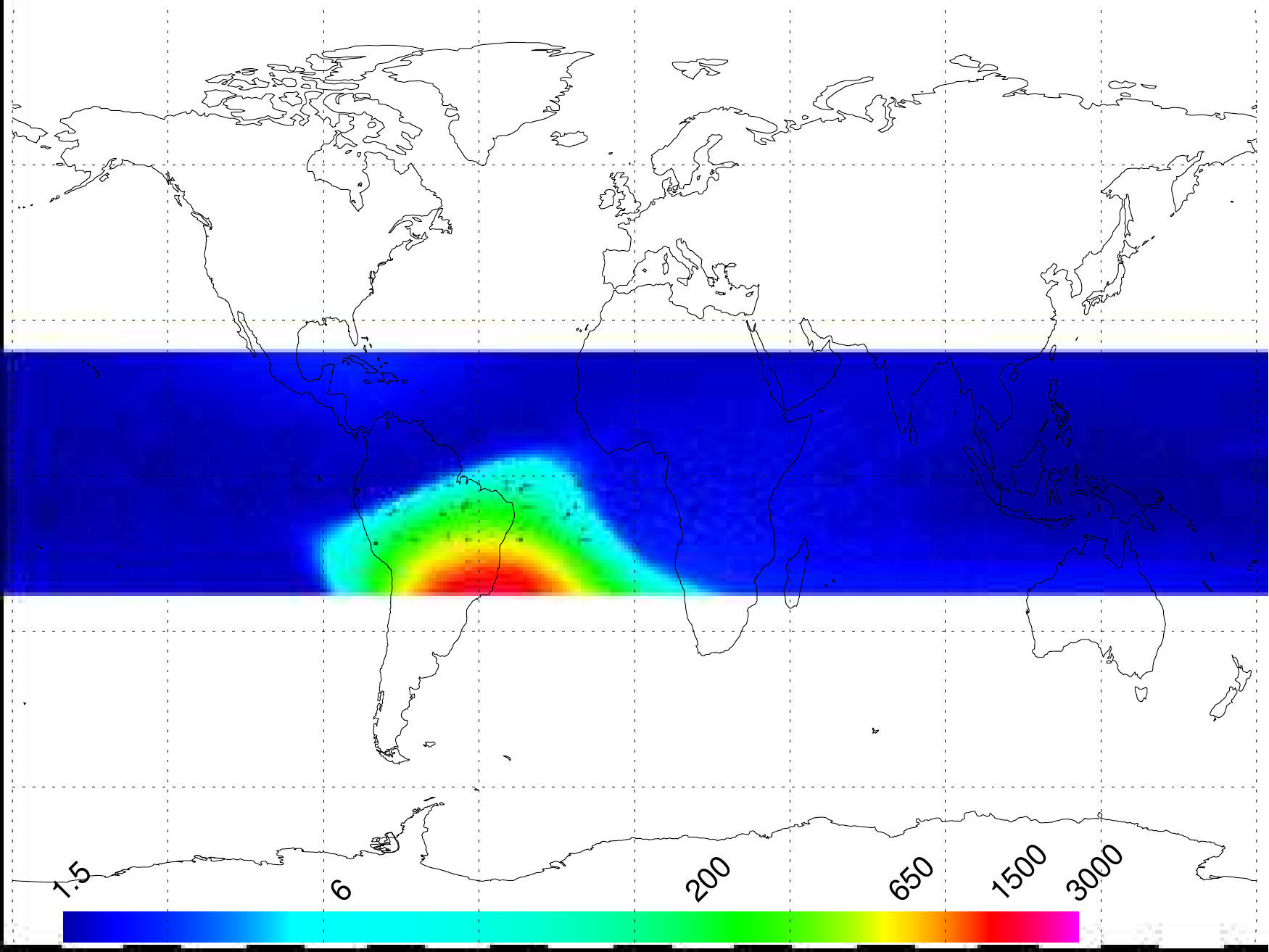}
  \hfill
   \includegraphics[viewport=3 3 502 375, clip, width=0.7\columnwidth]{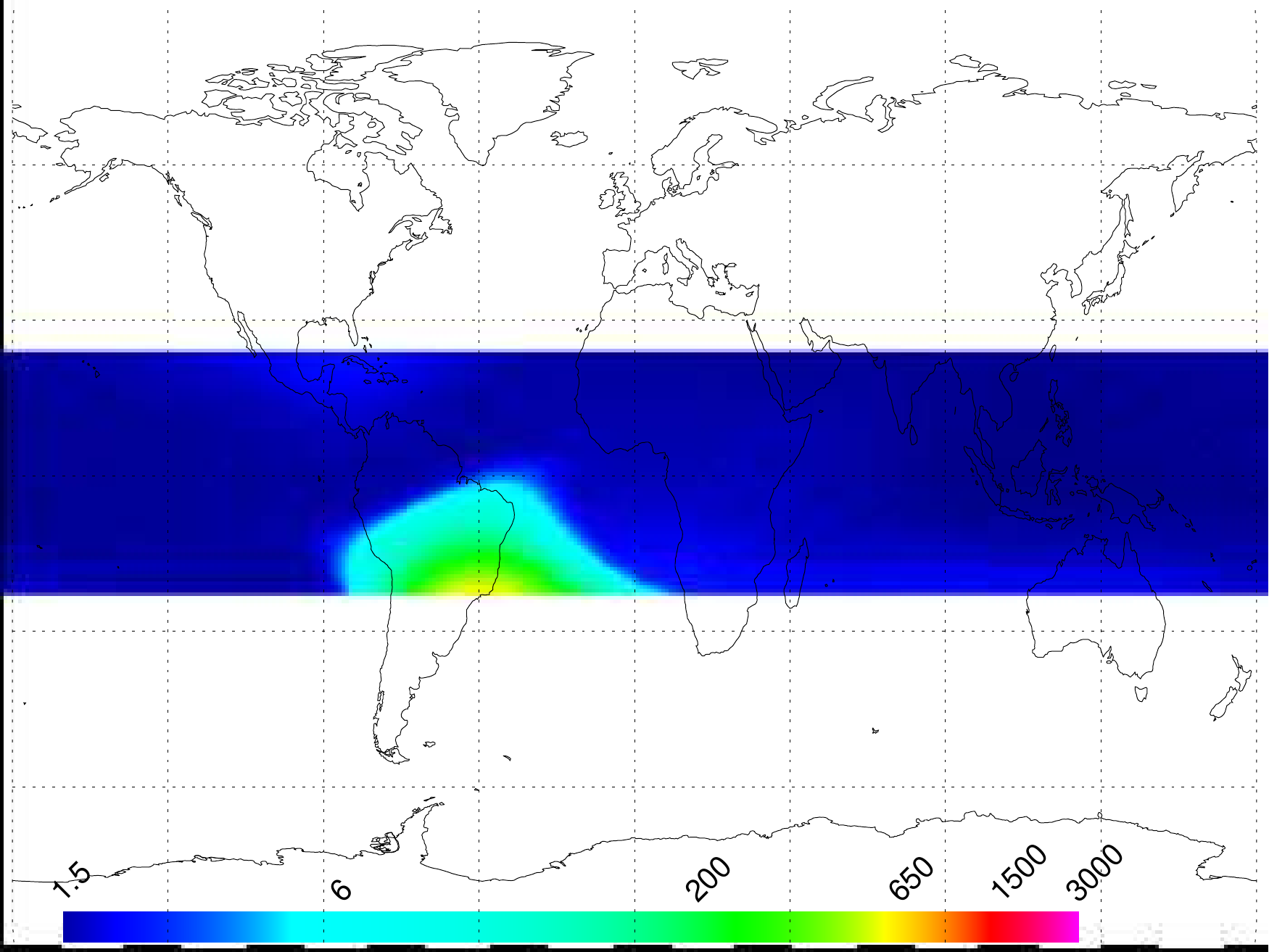}

  \caption{Maps of the average count rate of the particle monitor of
    \textit{HEXTE B}. \emph{Top:} 1997, $1^\text{st}$ quarter,
    altitude 570--579\,km. \emph{Bottom:} 2003, $3^\text{rd}$ quarter,
    altitude 500--509\,km.}
\label{fig:map97_03}
\end{figure*}

\subsection{Mapping the SAA}\label{sec:analy}
In order to study the shape and location of the SAA, averaged particle
flux maps were determined with a time resolution of 3\,months and a
spatial resolution of $0.25^\circ$ in longitude, $0.5^\circ$ in
latitude, and 10\,km in altitude.
The grid resolution chosen is a compromise between the desire to
obtain a good spatial resolution and a good signal to noise ratio in
the resulting maps. In the southernmost latitude bins, which are closest to the
center of the SAA and on which most of the following
discussion is based, each bin represents the average of at least
6~measurements. Towards the equator the number of measurements
decreases. On average, each pixel is visited 1.5 times.
Figure~\ref{fig:map97_03} shows two examples for the generated maps
which show the countrate projected on the Earth. We have chosen the
years 1997 and 2003 as examples as they provide good statistics and
show the typical change of the SAA over the years. 

As discussed in Sect.~\ref{sec:rxte}, atmospheric drag causes the
satellite to constantly lose altitude. For this reason, the highest
altitudes are only mapped in the first few years, while lower
altitudes are only mapped more recently. To indicate this change, in
this paper the different colors and symbols listed in
Table~\ref{tab:symbos} are used in all plots displaying the time
dependence of an SAA related quantity. This approach allows us to
check whether the general trends discussed in the next sections are
altitude dependent. We note that while some properties of the SAA are
altitude dependent \citep[see, e.g.,][]{ginet06a}, in the following we
will concentrate on the shape and position of the SAA, which have been
shown to be independent of altitude \citep{grigoryan05a}.

\begin{table}
  \centering
  \caption{Symbols and colors used in the plots. The last column gives
    the time-bins in years for which data for that specific altitude
    are available. Time-bins are 3 months in duration
    each.}\label{tab:symbos} 
  \vspace*{3mm}
 \begin{tabular}[t]{llcc}
    \hline
    \textbf{Symbol} & \textbf{Color} & \textbf{Altitude} (km) &
    \textbf{Time-Bins}\\\hline
    cross ($+$) & red & 580--589 & 1996.00--1999.50 \\
    asterisk ($\ast$) & violet & 570--579 & 1996.00--2000.25\\ 
    diamond ($\diamond$) & orange & 560--569 & 1996.00--2001.00\\
    triangle ($\blacktriangle$) & light blue & 550--559 & 1999.50--2001.75\\
    square ($\blacksquare$) & green & 540--549 & 2000.50--2002.00\\
    times ($\times$) & orange & 530--539 & 2001.25--2002.75\\ 
    cross ($+$) & green & 520--529 & 2001.75--2003.25\\
    asterisk ($\ast$) & red & 510--519 & 2002.25--2004.50\\ 
    diamond ($\blacklozenge$) & dark blue & 500--509 & 2002.75--2006.25\\
    triangle ($\blacktriangle$) & turquoise & 490--499 & 2003.75--2007.75\\\hline 
      \end{tabular}

\end{table}

\section{Methodology}\label{subsec:def}
\begin{figure}
   \centering
    \includegraphics[viewport=42 355 382 667, clip, width=0.7\columnwidth]{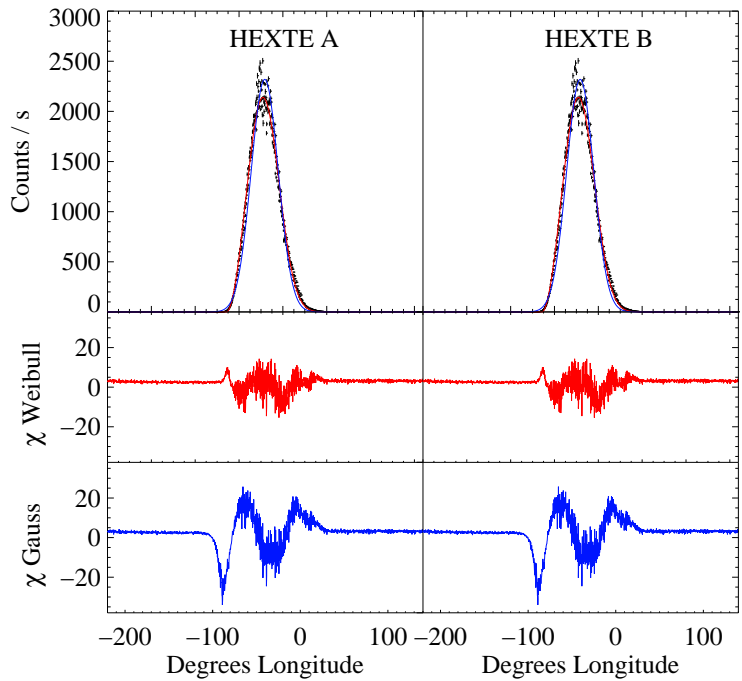}
   \caption{Top: Simultaneous fit of HEXTE A (left panels) and B
     (right panels) monitor for 1997, $1^\text{st}$ quarter, altitude
     570 -- 579\,km and latitude $-23^\circ$. Middle: Residuals of the best
     fit Weibull function (red curve in the top panel). Bottom:
     Residuals of the best fit Gaussian function (blue curve in the
     top panel).} \label{fig:simult_fit}
\end{figure}

The distribution of the particle flux with longitude has traditionally
been described using Gaussian functions \citep[e.g.,][and references
therein]{konradi94a,buehler96a}, with the peak of the Gaussian
describing the position of the SAA and its width being a measure for
its size. A closer inspection of the particle monitor rates along one
latitude bin reveals, however, that the SAA is distinctly asymmetric
in shape (see Fig.~\ref{fig:simult_fit}). An asymmetric function is
thus better suited to model the overall longitude-dependent particle
flux. Numerical experimenting revealed that for all measurements
described here a Weibull function \citep{weibull51a} defined by

\begin{equation}\label{eq:weibull}
  y(x; A,k,\lambda,\theta) = 
\begin{cases}
A \cdot \frac k \lambda \cdot \left(\frac {x - \theta} {\lambda} \right)^{k-1}
  \cdot \exp\left(-\left(\frac {x - \theta} {\lambda} \right)^k\right)
& \text{for $x\ge \theta$} \\
0 & \text{for $x<\theta$}
\end{cases}
\end{equation}

gives a significantly better description of the shape of the SAA in
terms of the $\chi^2$-value of the fits. Here, $x$ is the geographic longitude and the fitparameters are $A$, the
normalization (or strength), $\lambda$ the scale parameter, $\theta$ the shift
parameter, and $k$ the shape parameter of the Weibull function. Weibull functions are applicable for a wide range of phenomena \citep[see][for an overview]{brown51a}, but have to our knowledge not yet been used to described the shape of the SAA. 
The Weibull function is normalized such that

\begin{equation}
        \int_{-\pi}^{+\pi} y(x;A,k,\lambda,\theta) \,dx= A
\end{equation}

i.e., the normalization $A$ is proportional to the total longitude
integrated particle flux. In the following, we will call $A$ the
strength or flux of the SAA.

The position of the maximum of the Weibull function, $\bar x$, is

\begin{equation}\label{eq:posmaxwei}
 \bar x =  \lambda \cdot \left(\frac{k-1}{k}\right)^{\frac 1 k} + \theta 
\end{equation}

and the variance of this distribution, $\sigma^2_\text{wei}$, is given
by

\begin{equation}\label{eq:varweib}
  \sigma^2_\text{wei} = \lambda^2 \cdot \Gamma\left(1 + \frac 2 k
                                             \right) - \mu^2  
\end{equation}

where $\Gamma(x)$ is the Gamma-function. To illustrate that a Weibull
function indeed describes the data better than a simple Gaussian,
Fig.~\ref{fig:simult_fit} compares fits of both functions to data
taken at $-23^\circ$ latitude early in the mission. With
$\chi^2=4.39\cdot10^4$ and 2875 degrees of freedom, the best-fit Weibull function describes the data
significantly better than a Gaussian function ($\chi^2=1.46\cdot10^5$, 2876 degrees of freedom). For times
taken during the solar maximum later in the mission, the SAA shape
becomes more symmetric and Gaussians describe the shape almost as well
as the Weibull function. Even during those times, however, the Weibull
fits consistently give lower $\chi^2$-values than Gaussian fits. In
the following, we will therefore only present results from the Weibull
fits. We note, however, that the main results presented in this paper
are independent of the specific fit function employed and only differ
in minor details.

\begin{figure}
   \centering
    \includegraphics[viewport=64 370 550 702, clip, width=0.7\columnwidth]{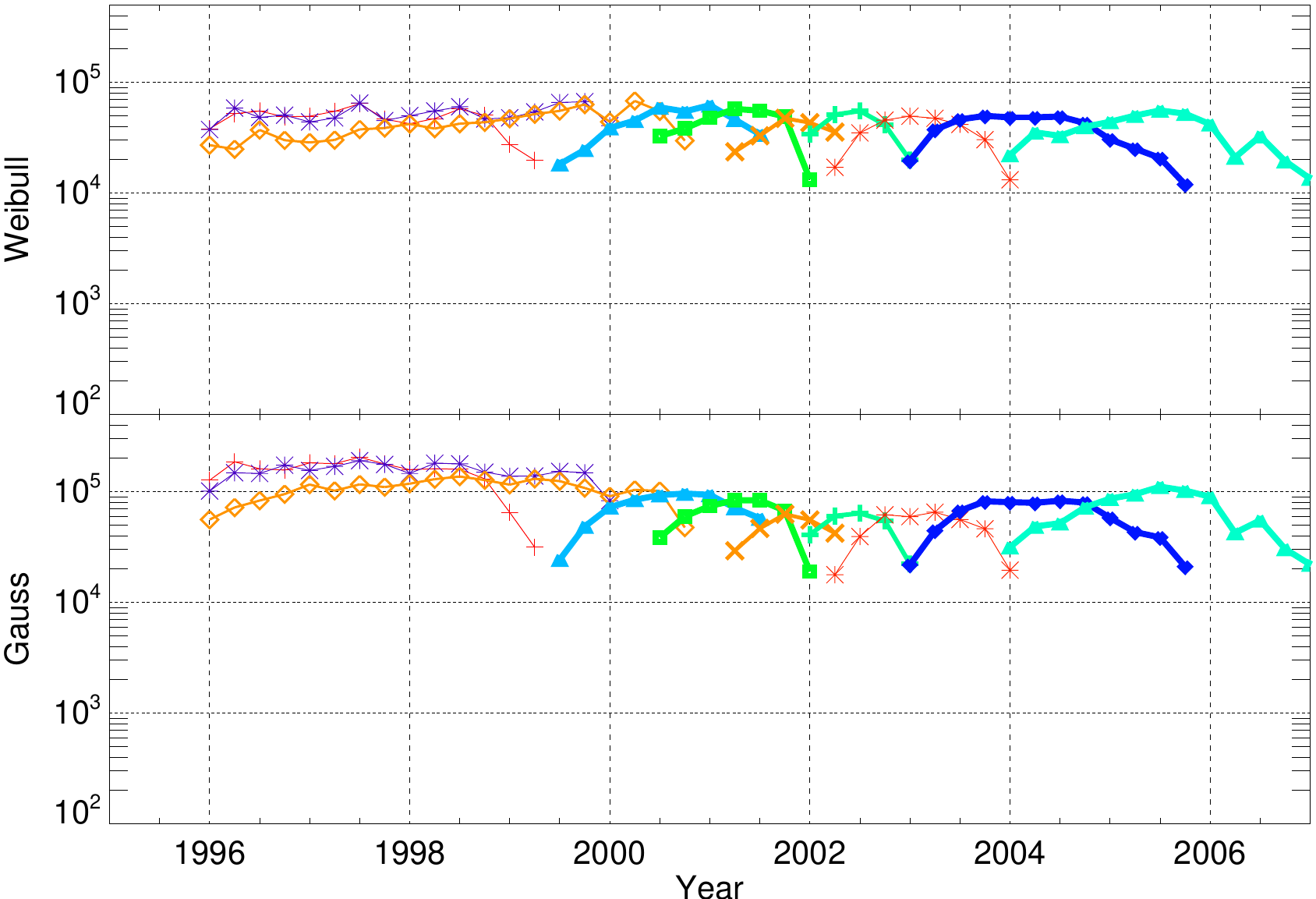}
   \caption{$\chi^2$-values of the Weibull (top) and Gaussian (bottom)
     fit for the duration of the campaign. Fits have typically $\sim$2500
     degrees of freedom and were done for the southernmost latitude. The lower $\chi^2$-values at the start and
     end of each altitude bin interval are due to the sparseness of
     the data and therefore the larger error bars. See Table~\ref{tab:symbos} for an explanation of the symbols used. }
   \label{fig:chi_ad_both}
\end{figure}

To allow us to gauge the overall quality of the data modeling,
Fig.~\ref{fig:chi_ad_both} shows the $\chi^2$-values of the Weibull
and Gaussian fits for the entire campaign. The quality of the fits is
usually very good. The large values of the reduced $\chi^2$ of around 8 can be ascribed to the fact that in the maximum of the latitude cut the data shows a detailed substructure which can not be well described by a simple smooth function such as a Weibull or a Gaussian.
At the end of every altitude bin, the $\chi^2$-values decrease
dramatically. This decrease is due to the fact that at the end of a bin only
very few measurements are taken at that altitude, increasing the
measurement uncertainty and thus decreasing $\chi^2$. Data points with $\chi^2 < 10^4$ were consequently not taken into account in the modeling described below.

\section{Results: Evolution of the SAA 1996--2007 }\label{sec:results}

\begin{figure*}
    \includegraphics[viewport= 47 364 391 696, clip, width=0.6\columnwidth]{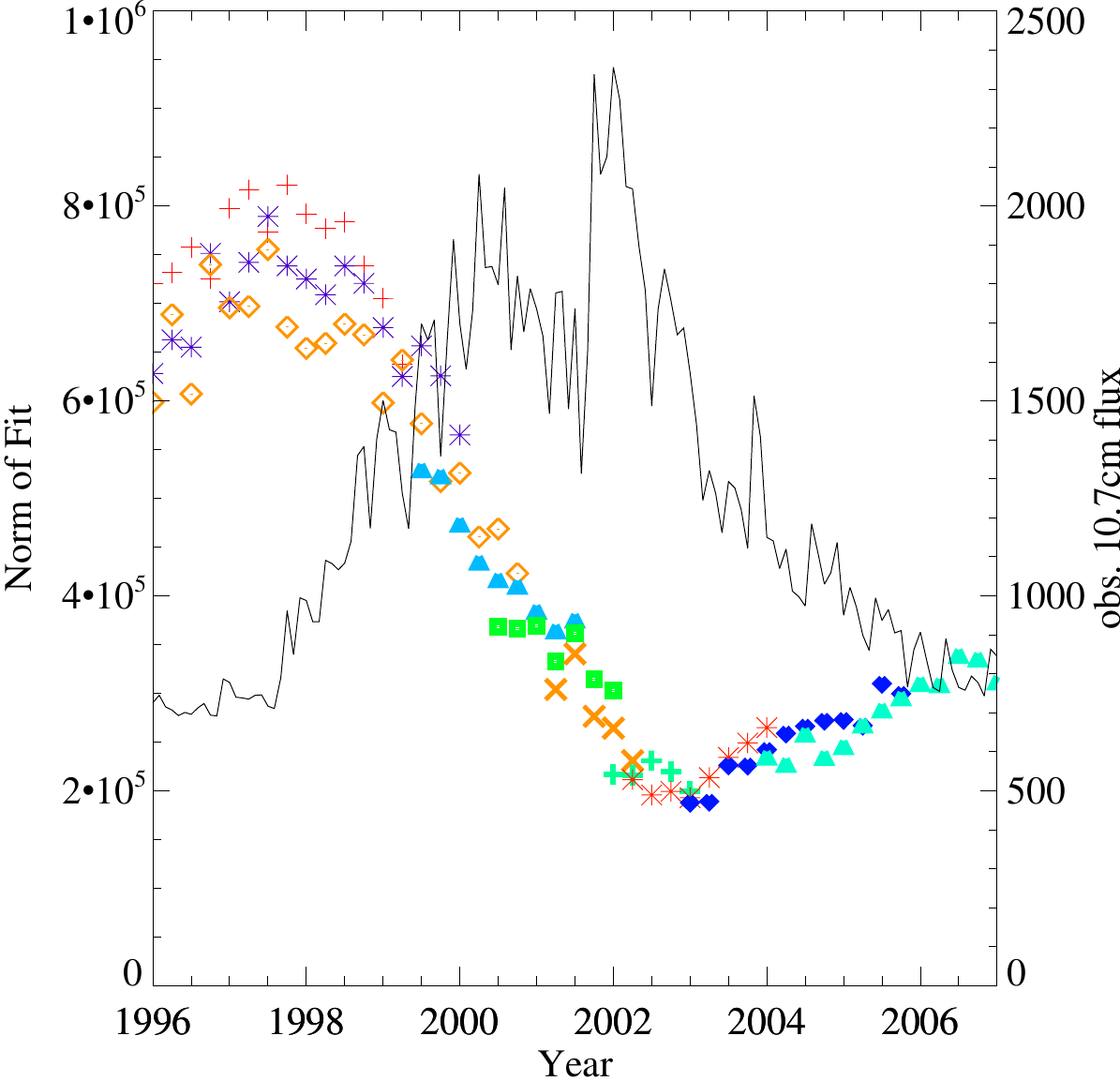}
    \hfill
     \includegraphics[viewport=47 364 391 696,clip,width=0.6\columnwidth]{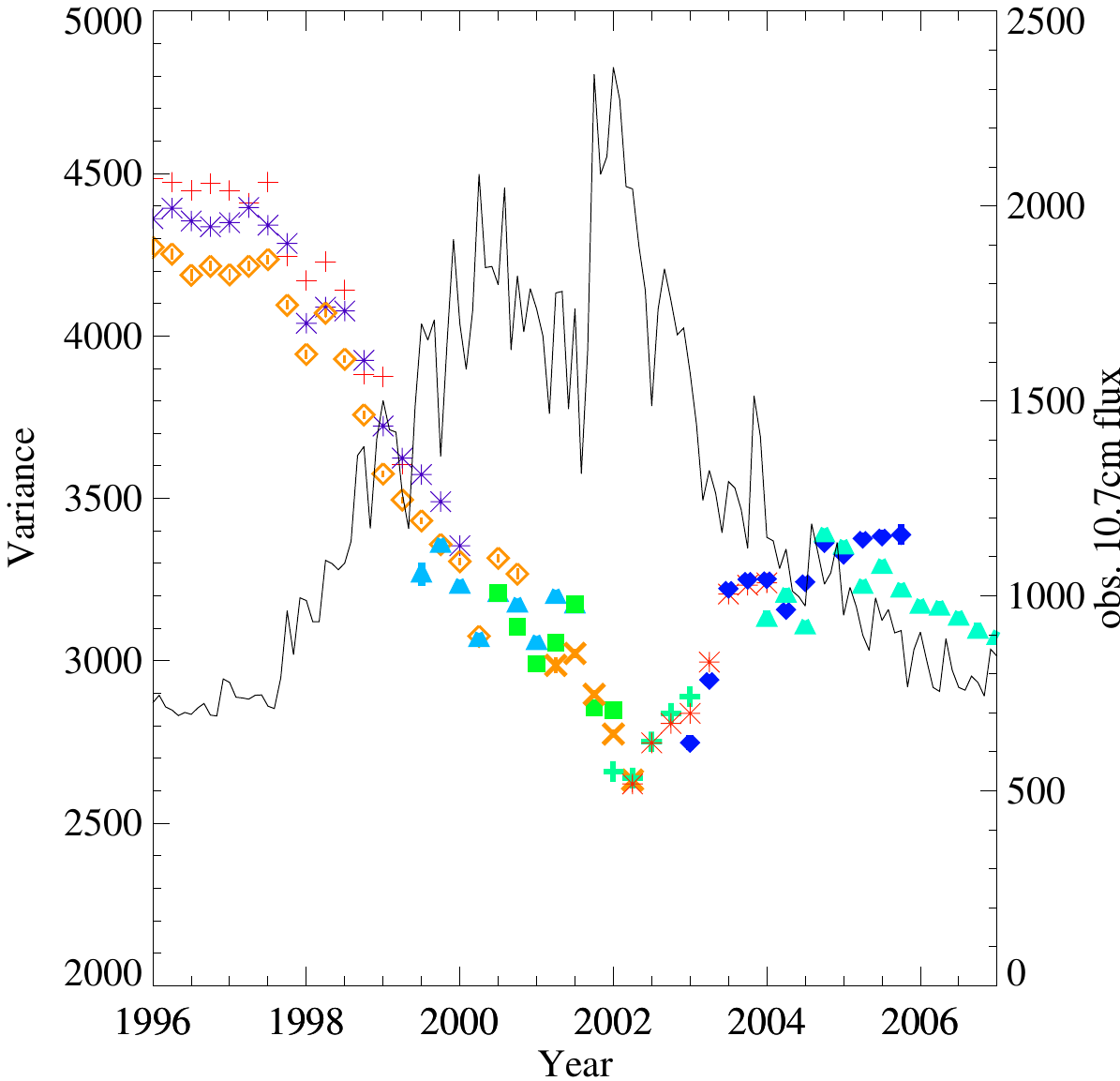}

   \caption{Evolution of the strength and width of the SAA from 1996 to 2007 as
     parametrized by the normalization (left) and the variance (right) of the best fitting Weibull function. See
     Table~\ref{tab:symbos} for an explanation of the symbols used.
     The black line shows the solar 10.7\,cm radio flux.}
   \label{fig:norm}
\end{figure*}

In this section we describe the results of our study of the evolution
of the SAA through one solar cycle, from 1996 until the end of 2007.
In our analysis, we concentrate on the southernmost tracks of
\textit{RXTE}, i.e., the HEXTE particle monitor count rate measured in
the southernmost or the 11 southernmost latitude bins. In these bins
the signal to noise ratio is highest. In addition, these bins are
the ones closest to the center of the SAA and therefore most
representative for its overall behavior. While we also performed some
altitude dependent analysis, the main emphasis will be on the
dependence of the strength of the particle flux in the SAA
(Sect.~\ref{susec:tempvar}) and on the motion of the SAA between 1996
and 2007 (Sect.~\ref{susec:move}). 

\subsection{Temporal Variation of the strength}\label{susec:tempvar}
\begin{figure}
   \centering
    \includegraphics[viewport= 73 370 380 674, clip, width=0.7\columnwidth]{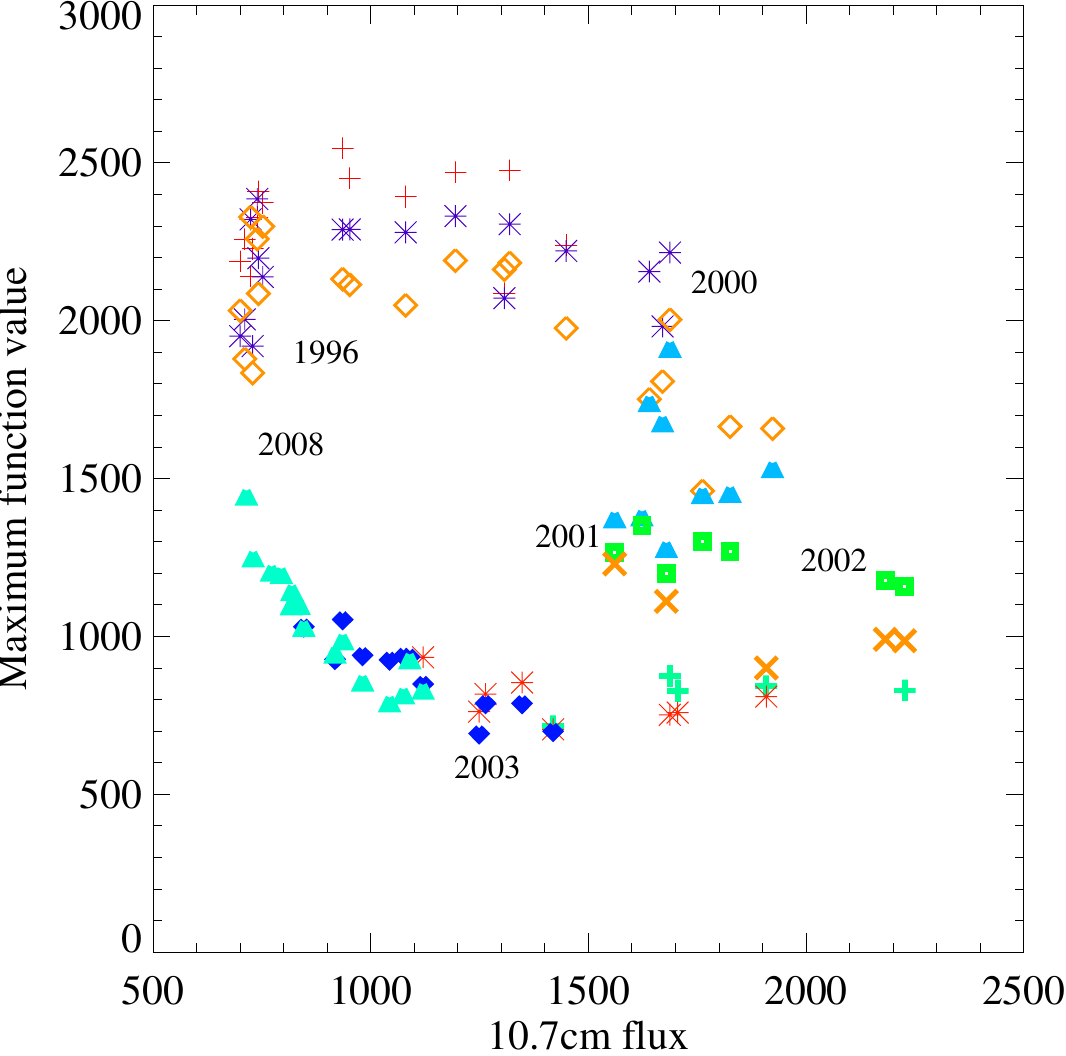}
   \caption{The maximum value of the Weibull fit versus the solar
     10.7\,cm flux. The color code used is described in
     Table~\ref{tab:symbos} and represents the altitude bins (and thus
     indirectly the time). The years indicate the approximate time of the data acquisition.}\label{fig:maxweivs10.7}
\end{figure}

Figure~\ref{fig:norm} shows the evolution of the normalization ($A$ in Eqn. \ref{eq:weibull}) and the variance for
the Weibull fits from 1996 to 2007. These parameters are good proxies
for the evolution of the particle flux in the SAA during that time
interval. As previously shown, e.g., by \citet{buehler96a},
\citet{huston96a}, and \citet{dachev99a}, both parameters are inversely
correlated with the 10.7\,cm radio flux from the Sun measured in $10^{-22}$\,W\,m$^{-2}$\,Hz$^{-1}$ (obtained by the
National Geophysics Data
Center\footnote{The NGDC can be reached at
  \url{http://www.ngdc.noaa.gov}, the 10.7\,cm data are available at 
  \url{ftp://ftp.ngdc.noaa.gov/STP/SOLAR_DATA/SOLAR_RADIO/FLUX/penticton.txt}.}\!),
which is a measure of the solar activity. As outlined by
\citet{huston96a} and \citet{dachev99a}, this inverse correlation is
due to heating of the upper atmosphere during times of higher solar
activity. This heating leads to a higher
neutral density in the altitude region of the SAA and consequently to
a higher absorption and deflection of trapped particles, which results in a lower
particle flux compared to times of lower solar activity. A
cross-correlation analysis shows a maximum of the absolute value of
the cross correlation function of $\sim$0.75 at a time delay of
1\,year. This result is in agreement with \citet{huston96a}.
This time lag between the particle flux as characterized by the normalization of
the SAA and the solar 10.7\,cm flux is also illustrated in
Fig.~\ref{fig:maxweivs10.7} where we plot both quantities against each
other. As the \textsl{RXTE} data are available for $\sim$12\,years,
they cover one full 11\,year solar cycle, and consequently lie on an
ellipse as is characteristic for physical quantities exhibiting an
hysteric behavior. A few outliers can be seen in Fig.~\ref{fig:maxweivs10.7} around 2001, which can be ascribed to a short term reduction of the solar 10.7\,cm flux in this time window. The particle environment has damped out these short term fluctuation so that the 2001 data points do not fall perfectly onto the ellipse.

\begin{figure}
   \centering
    \includegraphics[viewport= 61 364 392 695, clip, width=0.7\columnwidth]{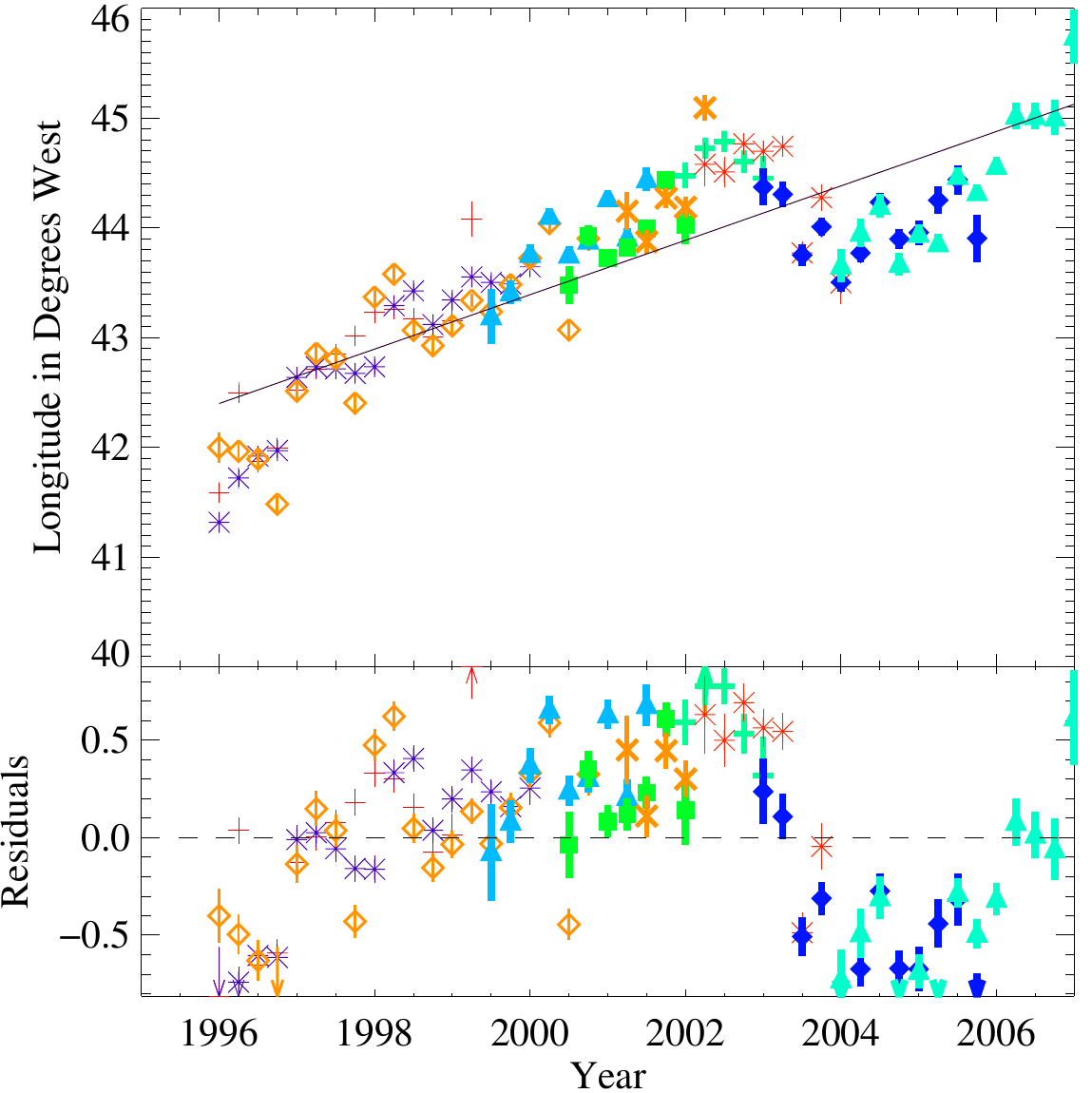}
   \caption{Position of the maximum of the Weibull fit for
     latitude $-23^\circ$ and a linear fit to these data with a slope of
     $0.248(3)^\circ / \text{yr}$. } \label{fig:mov_mw_1f}
\end{figure}

\subsection{Movement of the SAA}\label{susec:move}
To characterize the drift of the SAA we use the
position of the maximum of the fits for the southernmost latitude bin.
Figure~\ref{fig:mov_mw_1f} clearly shows a roughly linear motion in
longitude, with a linear fit to the data resulting in a westerly
motion of $0.248 \pm 0.003^\circ / \text{yr}$, which is in good
agreement with results of other groups \citep[see][and references therein]{badhwar97a}. In
contrast to these earlier results, which were generally based on
sparse data sampling, Fig.~\ref{fig:mov_mw_1f} clearly reveals
significant deviations from the overall linear trend. Most obviously,
a strong change in the position of the SAA is clearly visible between
2003 and 2004, where the SAA's center is moving towards the east,
i.e., in the opposite direction of the usual movement. This behavior
is seen in both, the Weibull and the Gaussian fits, and it is also
present when considering not the maximum of the fits but the
expectation value of the data. In addition, there is also no change in
the quality of the Weibull or Gauss fits during that time
(Fig.~\ref{fig:chi_ad_both}). The drift rates found in all of these
approaches are very similar. 

\begin{figure}
   \centering
    \includegraphics[viewport=49 364 345 678, clip, width=0.7\columnwidth]{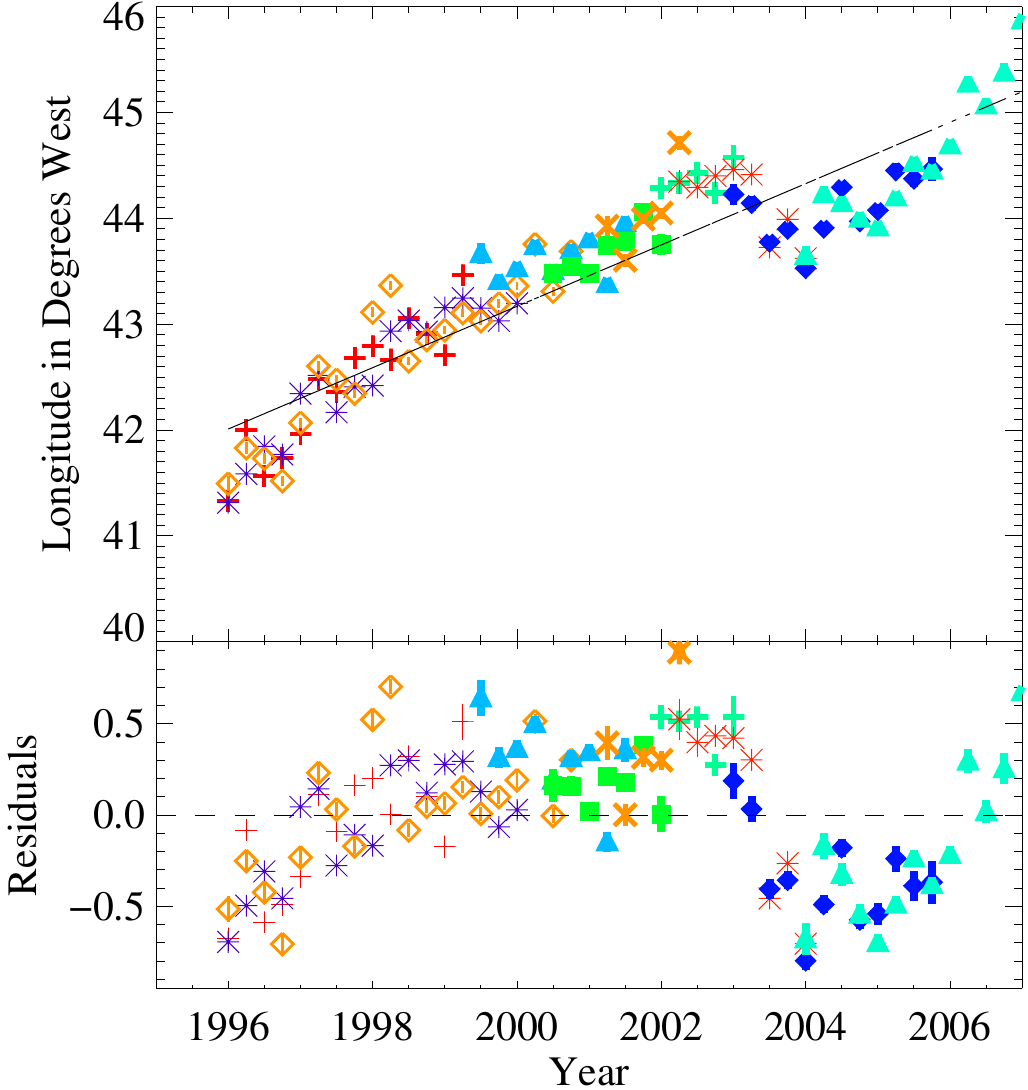}
   \caption{Average position of the maximum of the Weibull fits for the data between latitude $-23^\circ$  and $-18^\circ$. Superposed to the data is a
     linear fit to the data with a slope of
     $0.290(2)^\circ / \text{yr}$. }\label{fig:mov_mw2_1f}
\end{figure}

Since the signal to noise in the southernmost latitude bin is too low
to allow a more detailed analysis of the drift of the SAA, we averaged
the maximum position over the 11 southernmost latitude bins. All of
these bins are still close to the maximum of the SAA and the latitude dependence of the particle count rate is very similar. Figure~\ref{fig:mov_mw2_1f} shows that the averaged
position shows the same behavior as the position for latitude $23^\circ$
south only. This means that the sudden change between 2003 and 2004 is
a persistent feature and is independent of latitude. A linear fit
leads to a movement rate of $0.290 \pm 0.002^\circ / \text{yr}$.
The residuals of this fit clearly show that in addition to the jump
around 2003 another change of slope is apparent in 1998 (these
residuals are also present in Fig.~\ref{fig:mov_mw_1f}, but are only
of borderline significance in these data).

\begin{figure}
   \centering
    \includegraphics[viewport= 61 365 388 694, clip, width=0.7\columnwidth]{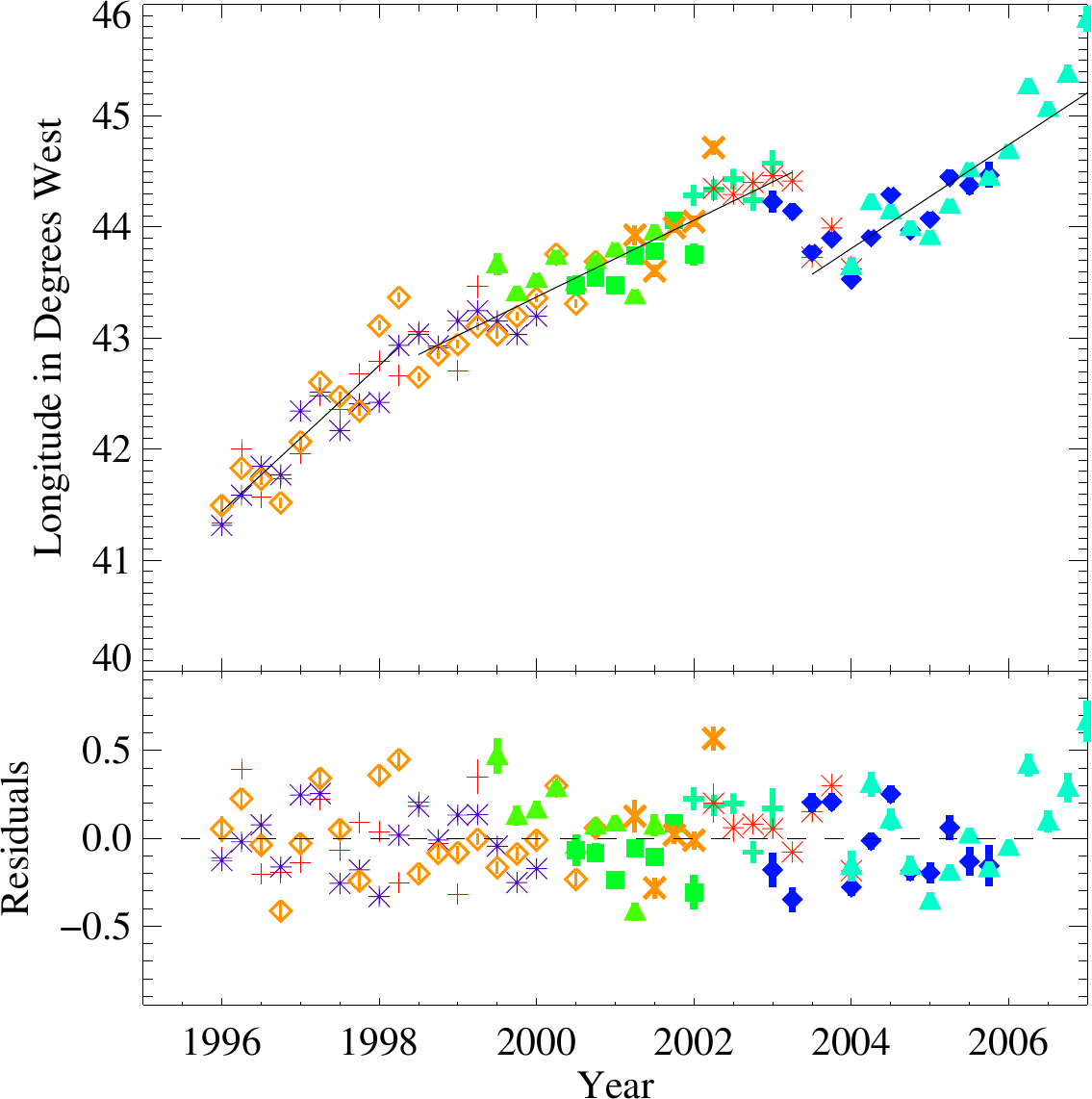}
   \caption{Average position of the maximum of the Weibull fits for the data between latitude $-23^\circ$ and $-18^\circ$. Superposed are three linear
     fits, with a slope of $0.65(1)^\circ / \text{yr}$
     (1996.0--1998.25), $0.346(5)^\circ / \text{yr}$
     (1996.5--2003.25), and $0.46(1)^\circ / \text{yr}$
     (2003.5--2007.0), respectively. }  \label{fig:mov_mw2_3f}
\end{figure}

Since a simple linear fit is not a good approximation to the movement
of the SAA due to the sudden change in the movement in 1998 and 2003,
we next parametrized the data by fitting it with piecewise linear
functions, allowing for two ``jumps'' in the data. This approach gives
a good description of the overall data, with the slope changing from
$0.656(11)^\circ / \text{yr}$ for the time interval before
1998.25, to $0.346(5)^\circ / \text{yr}$ for the interval
1998.25--2003.25, and $0.467(12)^\circ / \text{yr}$ since
then (Fig.~\ref{fig:mov_mw2_3f}). We note that with $\chi^2=2040.78$
for 104 degrees of freedom, this fit is significantly better than a
fit in which the 1998.25-break is ignored. In the latter case,
$\chi^2=2527.8$. Using the $F$-test to test the reasonability of the
second break point of the linear fit gives a very low probability
value of $4.065\cdot 10^{-6}$ for the break to be a random artifact.
Fits for the other ways used to determine the maximum of the SAA as
discussed above lead to the same dates for the break points.

\begin{figure}
   \centering
    \includegraphics[viewport= 61 365 388 694, clip, width=0.7\columnwidth]{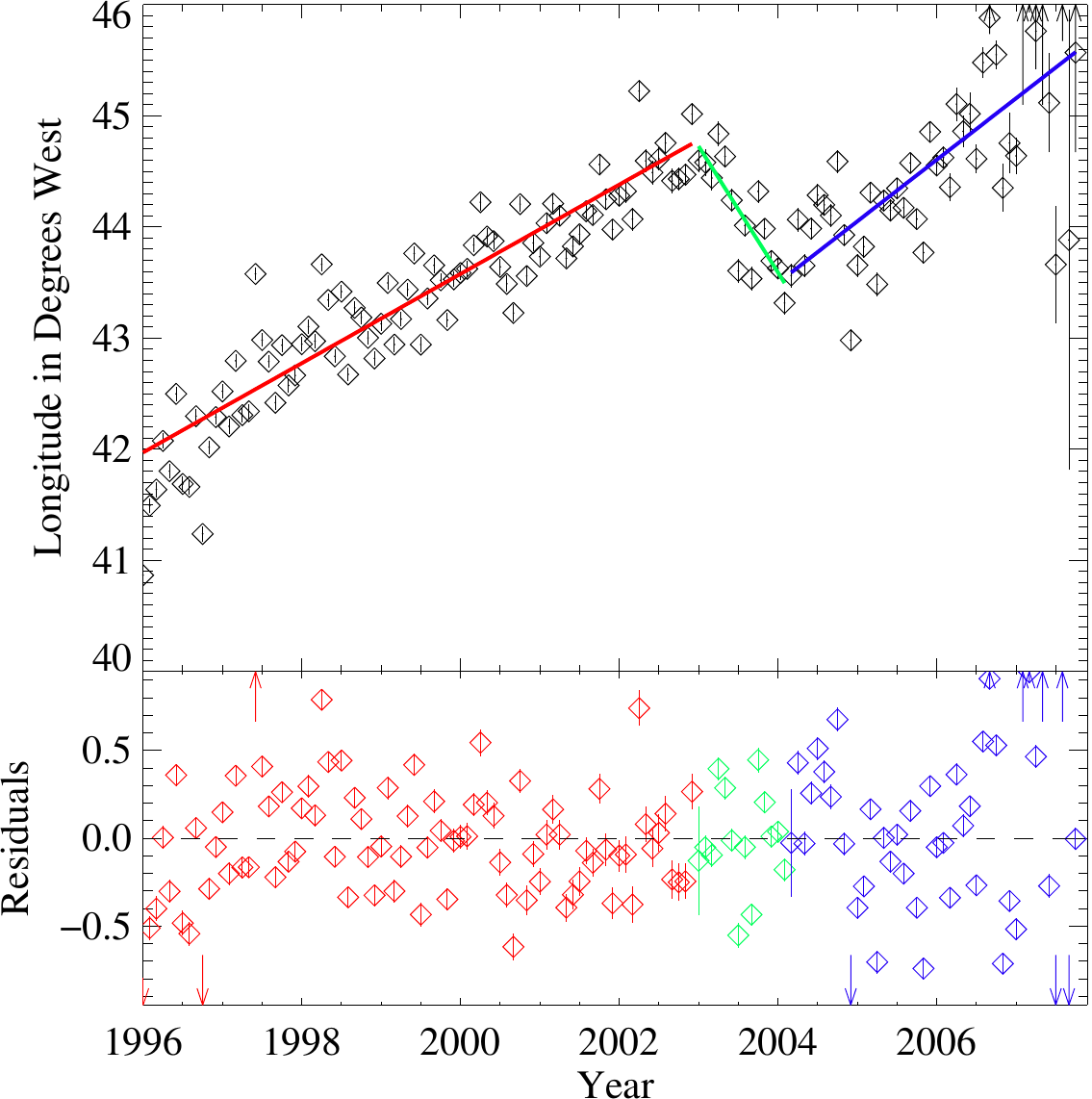}
   \caption{Position of the maximum of the Weibull fit for
     latitude $-23^\circ$ and the best-fit piecewise linear function
     describing the motion of the SAA during the 2003 geomagnetic
     jerk. The second panel shows the residuals of the fit, with
     colors indicating the three parts of the fit.}
   \label{fig:mov_htr_maxwei_3f}
\end{figure}

To determine the start and end of the change more precisely, we
extracted the \textsl{RXTE} data again with a higher temporal
resolution of 1\,month. It was necessary to average over all altitudes
to retain a good signal to noise ratio. This approach is justified
since our previous studies confirm prior investigations showing that the
position of the maximum of the SAA does not depend on the altitude for
the small altitude range considered here \citep{grigoryan05a}.
Figure~\ref{fig:mov_htr_maxwei_3f} shows the position of the maximum
of the Weibull fit with a resolution of 1\,month. Superposed to the
data is again a piecewise linear fit in which the slope is allowed to
change at two different dates. A $\chi^2$-minimization shows that the
changes in slope occurred in February 2003 and in April 2004. The SAA
region was therefore moving opposite to its usual direction of
movement for slightly more than 1\,year. With a movement rate of $1.14
\pm 0.094^\circ / \text{yr}$ eastwards, the SAA moved significantly
faster than usual. The feature in 1999, which was found in the
altitude dependent data, is still seen weakly in the high time
resolution data. The backwards drift was distinctly shorter than
the one in 2003, however, and thus the duration of the event cannot be
determined.

We conclude that a sudden change in the movement of the SAA seems not to be
rare, it is only the amplitude of the 2003.25-move that seems to be
special. A sudden change means that the SAA
moves quickly towards the east, during the third quarter of 2003 by as
much as about $0.8^\circ$. Additionally the rate of the movement is different
after each discontinuity. The commonly assumed rate of
$0.3^\circ / \text{yr}$ seems to be valid only on
longer timescales. On shorter timescales the rate can differ
drastically from this value. 

\section{Summary and conclusion}\label{sec:disc}
Using the particle monitor aboard \textit{RXTE} we have studied almost
12\,years of continuously taken data of the radiation background in
the low Earth orbit of this satellite. While the measurements could
not determine the spectrum of the particle radiation, they allowed a
detailed study of the shape and position of the SAA. We have shown
that the shape of the SAA along one latitude can be better described by a
Weibull function than by a pure Gaussian. The data show a clear anti-correlation between the
10.7\,cm solar flux and the intensity and size of the SAA, with a lag
of about 1\,year. The large time range of our data allowed us to
monitor the SAA for one solar cycle and we could show that after
approximately 11\,years the particle flux in the SAA has about the same strength
again. 

The most important result of this study was that the location of the
maximum of the fit as a function of time could be
determined with a time resolution of a few months for the whole
12\,years of data. The \textit{RXTE} data confirm earlier
measurements of a westward drift of the SAA, with a rate of $\sim$0.27$^\circ / \text{yr}$, depending on what parametrization
is chosen for its shape. The data also show, however, that the drift
rate is not constant, but shows two irregularities, meaning that its
position shifts slightly eastwards and then continues to move
westwards. These break points were determined to be during the first
quarter of 1998 and in the second quarter of 2003, with the latter one
being distinctly stronger than the first one.

These irregularities could indicate a change in the particle
population of the SAA. There are several phenomena which can produce
such a change. On shorter timescales, solar coronal mass ejections
(CME) can cause geomagnetic storms \citep{brueckner98a} and inject
electrons and protons into the atmosphere. \citet{asikainen05a} have
shown that during the great storm of March 21, 2001 electrons became
trapped in the SAA and drifted around the Earth together with the SAA.
In November 2003 a major CME caused an intense geomagnetic storm,
investigated by \citet{beckerguedes07a}. While coincident with the
time interval of the jump reported here, the jump started
significantly earlier than the CME. Furthermore, the injected
electrons that were seen in 2001 decay with a lifetime of only about
8\,hours \citep{asikainen05a}, while the feature seen in the SAA has a
longer lasting influence. These facts make it highly unlikely that
CMEs are responsible for the measured irregularities.

As CMEs are the most prominent outside source which disturbs the
geomagnetic field and the particle population around the Earth and only have a short term effect on the SAA, it is
more likely that the origin for the changes in the movement is not in
the particle distribution, but rather in the field itself. The most probable features of the field responsible for such events are the ``geomagnetic jerks''. As described by \citet{olsen07a}, in early 2003 a large geomagnetic
jerk occurred. This jerk might even have had an influence on the drift
of the north magnetic dip pole, which accelerated for over 10 years
but started to decelerate in 2003 \citep{olsen07b}. This result shows
that the jerk had a major influence on the magnetic field of the
Earth. \citet{olsen08a} furthermore analyzed satellite based geomagnetic data and could explain the observed accelerations in the secular variation via a relaxed tangentially geostrophic flow model. They found that the core flows can fluctuate on time-scales of a few months, giving rise to equally fast changes in the geomagnetic field. It is likely that such changes could also influence the movement and the location of the SAA as these features depend strongly on the magnetic field.  The timescales given by \citet{olsen08a} agree well with our data. We note that another, but weaker, geomagnetic jerk occurred in 1999 \citep{mandea00a}, right
around the time of the earlier change in slope of the SAA drift discussed here.

We do not know of any jerk or other magnetic distortion
which took place in early 2004 to account for the resumption of the
westward drift of the SAA, although \citet{olsen08a} noticed a strong variation in the field at the end of 2004. These variations, however, seem to occur distinctly later than the reversal of the direction of the drift. Nonetheless the eastward drift could be connected to the jerk in 2003 and the duration of the
eastward movement might be just showing the slow response of the trapped
particles in the atmosphere.

This paper presented mainly observational data and a concluding physical explanation is beyond its scope.
To analyze the proposed connection between jerks and the location of the
SAA in further detail, additional data are clearly required. We are
currently in the process of reducing data from the ``Reuven Ramaty
High Energy Solar Spectroscopic Imager'' (\textsl{RHESSI}), which was
launched shortly before the 2003 geomagnetic jerk. The instruments on
this satellite have a higher time resolution than those on
\textsl{RXTE} and also allow one to distinguish between electrons and
protons. \citet{hajdas04a} have shown that \textsl{RHESSI} is very
capable of investigating the SAA region. In addition, data from the
``Detection of Electro-Magnetic Emissions Transmitted from Earthquake
Regions'' satellite (\textit{DEMETER}), which was launched in 2004,
will become available. While \textit{DEMETER} could not measure the
2003 geomagnetic jerks and its influences on the trapped radiation, it
passes fully through the SAA. A map demonstrating the capability of
\textit{DEMETER} has been published by \citet{sauvaud08a}, which shows
the SAA clearly. Additionally to the available geophysics satellites,
other high-energy astronomy satellites such as \textsl{XMM-Newton} or
\textsl{INTEGRAL} have particle detectors on board which will allow to
study the evolution of the Earth's magnetosphere and to provide
independent measurements of the particle background on a long
timescale.

 \ack
We acknowledge partial support from Deutsches Zentrum f\"ur Luft- und
Raumfahrt grants 50OR8080 and 50QR0801, from National Aeronautics and
Space Administration grant NAG5-30720, and from Deutscher
Akademischer Austauschdienst travel grant D/06/29438.
We thank the complete \textit{RXTE} team, especially the HEXTE team, for their support. We also thank the referees for their very helpful comments.

\bibliographystyle{elsart-harv}

\end{document}